\documentclass[aps,twocolumn,prl,showpacs,color,psfig,epsf]{revtex4-1}
\usepackage{amsmath}
\usepackage{color}
\usepackage{amsfonts}
\usepackage{epsf}
\usepackage{graphicx}
\usepackage{epstopdf}
\usepackage{ulem}
\usepackage{upgreek}

\begin{document}

\title{Hydrodynamic oscillations and variable swimming speed in squirmers close to repulsive walls} 
%\title{Hydrodynamic and repulsive effects of model squirmers between two confining walls: periodic orbits and tunable swimming speed}

\author{Juho S. Lintuvuori$^1$, Aidan T. Brown$^2$, Kevin Stratford$^3$ and Davide Marenduzzo$^2$}
\affiliation{$^1$ Laboratoire de Physique des Solides, CNRS, Univ. Paris-Sud, Universit\'e Paris-Saclay, 91405 Orsay Cedex, France\\
  $^2$ SUPA, School of Physics and Astronomy, University of Edinburgh, UK.\\
  $^3$ EPCC, School of Physics and Astronomy, University of Edinburgh, UK.
}

\begin{abstract}
We present a lattice Boltzmann study of the hydrodynamics of a fully resolved squirmer, confined in a slab of fluid between two no-slip walls. We show that the coupling between hydrodynamics and short-range repulsive interactions between the swimmer and the surface can lead to hydrodynamic trapping of both pushers and pullers at the wall, and to hydrodynamic oscillations in the case of a pusher. We further show that a pusher moves significantly faster when close to a surface than in the bulk, whereas a puller undergoes a transition between fast motion and a dynamical standstill according to the range of the repulsive interaction. Our results critically require near-field hydrodynamics; they further suggest that it should be possible to control the density and speed of squirmers at a surface by tuning the range of steric and electrostatic swimmer-wall interactions.  
\pacs{47.63.mf, 87.17.Jj, 47.63.Gd}
\end{abstract}

\maketitle

%introduction

{\it Introduction:} Motile organisms such as bacteria and sperm cells have a natural tendency to be attracted towards surfaces, and to swim near them~\cite{lord63}. This phenomenon may be relevant for the initial stage of the formation of biofilms, the microbial aggregates which often form on surfaces. Experiments have shown that this tendency is not unique to living swimmers, and is also exhibited by phoretic, synthetic active particles~\cite{brown14, takagi14, brown15}: in that context, it has been exploited for example, to attract microswimmers inside a colloidal crystal, where they orbit around the colloids~\cite{brown15}. The interaction between self-propelled particles and walls also provides a microscopic basis for the rectification of bacterial motion by asymmetric geometries ({\it e.g.} funnels)~\cite{galadja07}.
%of the mechanism through which bacterial motion can be rectified by asymmetric geometries, such as funnels~\cite{galadja07}.   

Previous work has proposed two possible mechanisms for surface accumulation of self-motile particles. A first view is that accumulation occurs through far-field hydrodynamic interactions~\cite{berke08}.
Another possibility is that motility itself, in the absence of solvent-mediated interactions, leads to accumulation~\cite{li09,elgeti13}: this mechanism requires a small enough channel, where the gap size is of the order of the typical distance travelled ballistically by the active particles, before rotational diffusion or tumbling reorients them.
The case of phoretic particles may be more complex~\cite{uspal15, brown15}, and may depend on the dynamics of the chemicals reacting at the swimmers surface. Schaar {\it et al.} recently showed that hydrodynamic torques strongly affect the ``detention times'' over which microswimmers reside near no-slip walls~\cite{schaar15}. The behaviour is partly controlled  by the details of the force distribution with which active particles stir the surrounding fluid; previous work has also shown that these determine the equilibria of a spherical squirmer near a flat surface~\cite{ishimoto13, li14}.

Previous theories have not systematically studied the effects of a short range repulsion between the particle and the wall; in practice this is always present in experiments, either due to screened electrostatics, for charged walls, or due to steric interactions, {\it e.g.} for polymer-coated surfaces. Here we show that explicitly including this repulsion is important, and strongly affects the dynamics near a surface. The interplay between hydrodynamics and short range repulsion can lead to trapping, periodic oscillations, and to a swimming speed significantly different from that in the bulk. These results provide an experimentally viable route to tune microswimmer concentration and speed near a no-slip surface. Furthermore, recent theoretical calculations~\cite{ishimoto13} predict that the equations of motion for squirmers which are pushers (exerting extensile forces on the fluid) do not possess any stationary bound state solution, {\it i.e.,} where the particle swims stably along the wall at fixed orientation. Our finding of oscillatory near-wall dynamics shows that even without such stationary solutions, trapping of swimmers at walls is possible.

%model and methods

\begin{figure}
\includegraphics[width=\columnwidth]{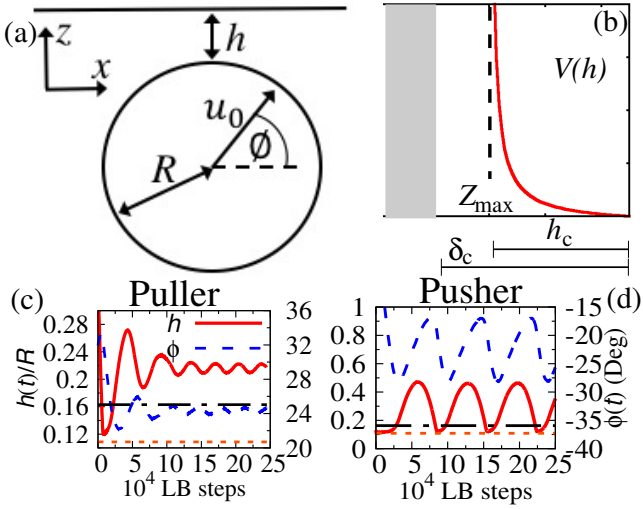}
\caption{(a) A cartoon showing an example of a squirmer near a flat wall, defining the gap size $h$ and angle $\phi$ used in the text. In the bulk, the squirmer would move with speed $u_0$ along the direction shown, at an angle $\phi$ from the horizontal. (b) There is a repulsive interaction between the wall and the squirmer. Examples of steady state $\phi(t)$ and $h(t)$ observed for (c) puller and (d) pusher dynamics near a flat wall in the absence of thermal noise are also shown. The interaction range is $\delta_c\sim 0.16R$ (dot-dashed line) and the potential diverges at $\sim 0.11R$ (dotted line).}
\label{Fig:Cartoon}
\end{figure}

{\it Squirmer model:} 
A popular model for swimmer hydrodynamics is a spherical squirmer -- a particle which is rendered self-motile through a surface slip velocity~\cite{lighthill52}. The squirmer model has been used to study, e.g., the collective motion and nutrient uptake of swimmers in thin films~\cite{zottl14,Lambert13}. To study the dynamics of a model squirmer in confinement we employ a lattice Boltzmann (LB) method~\cite{mikeLB}. To achieve time independent squirming motion, the following tangential (slip) velocity profile at the particle surface is used ~\cite{Magar03}
\begin{equation}
 u(\theta)=2\sum_{n=1}^{\infty}\frac{\sin \theta}{n(n+1)}\frac{dP_n(\cos \theta)}{d\cos \theta}B_n
\label{eq:veltangential}
\end{equation}
where $\theta$ is the polar angle and $P_n$ is the {\it n}th Legendre polynomial~\cite{ishimoto13}.

In the LB method a no-slip boundary condition at the fluid/solid interface can be achieved by using a standard method of bounce-back on links (BBL)~\cite{ladd1, ladd2}. When the boundary is moving ({\it e.g.} a colloidal particle) the BBL condition must be modified to take into account particle motion~\cite{ladd3}. These local rules can include additional terms, such as a surface slip velocity (Eq.~\ref{eq:veltangential}): in this way it is possible to simulate squirming motion~\cite{ignacio1,ignacio2}. Our implementation also includes thermal noise~\cite{adhikari}, allowing for simulations with a finite P\'eclet number.

{\it Simulation parameters:} We limit our simulations to simple squirmers with $B_n=0, n \ge 3$, but consider both pushers ($B_2 < 0$) and pullers ($B_2 > 0$). In simulation units (SU) we measure the lengths in lattice spacings and time in simulation steps. Parameters, all given in SU are: $B_1=0.0015$, $B_2= \pm 0.0075$, (which gives a swimming velocity in the bulk equal to $u_0=\tfrac{2}{3}B_1=10^{-3}$ and $\beta\equiv \tfrac{B_2}{B_1}=\pm 5$), fluid viscosity $\eta = 0.1$ and thermal noise $k_BT=10^{-5}$. We considered a fully resolved swimmer with radius $R=9.2$ (Fig. 1(a)). %\footnote{Assuming a particle radius $10\mu$m and using the kinematic viscosity of water $10^{-6}\tfrac{m^2}{s}$ a single simulation length and time unit can be mapped to $\sim 1\mu$m and $0.1\mu$s, giving a swimming speed $u_0=10\tfrac{\mathrm{mm}}{\mathrm{s}}$. The distance between the two walls is $\sim 100\mu$m, which corresponds to typical experiments~\cite{berke08}.} 
In order to model wall-particle repulsion, we employ a soft potential, $V(h)$, which goes smoothly to 0 as the wall-particle separation, $h$ (the gap size), approaches $\delta_c$, and diverges as $h\to 0$ (see Fig. 1(b), and Supporting Information~\cite{Suppl}).
The physics is governed by two main hydrodynamic dimensionless quantities: the Reynolds and P\'eclet numbers. Using the parameters above, these are Re=$\frac{u_0R}{\eta}\approx 0.09$ and $\mathrm{Pe}=\frac{u_0}{D_rR}\approx 2\times 10^{4}$ respectively, where $D_r=\frac{k_BT}{8\pi\eta R^3}$, is the rotational diffusion constant. Our simulations were carried out in a rectangular simulation box $120\times 120\times 96$, with periodic boundary conditions in $X$ and $Y$ and solid walls at $z=0$ and $z=95$.
To see how SU relate to physical units, we can, e.g., map a single length and time SU to $\sim 1\mu$m and $0.1\mu$s, respectively~\footnote{These would correspond to $R\sim$ 10 $\mu$m, $u_0\sim 10\tfrac{\mathrm{mm}}{\mathrm{s}}$, and a distance between walls of $\sim 100\mu$m.}.

\if{We use the following potential between the particle and the wall,
\begin{eqnarray}
 V_{cs}(h) = V(h) - V(h_{c}) - (h-h_c)\left.\frac{\partial V(h)}{\partial h}\right|_{h=h_c},
\end{eqnarray}\label{repulsion}
where the wall-particle-surface separation $h = Z_{max}^{min} \pm z - R$, and
\begin{equation}
V(h) = \epsilon\left(\sigma/h\right)^\nu.
\end{equation}
$V_{cs}$ has been cut-and-shifted to ensure that the potential and force go smoothly to zero at $h=h_c$. Parameters were chosen as $\epsilon = 0.004$, $\sigma = 0.1$, $\nu=1.0$. By choosing $Z_{\mathrm{max}} = Z_{\mathrm{top}} - (\delta_c - h_c)$ ($Z_{\mathrm{min}} = Z_{\mathrm{bottom}} + (\delta_c - h_c)$) and keeping $h_c=0.5$SU constant, we can have a well defined repulsion range $\delta_c$, while keeping $h_c$ and thus the potential form constant (Fig.1(b)). For the calculation of the gap size between the squirmer and the solid surface $h$ (Fig. 1(a)), we define the wall location half-way between the  solid node and first fluid node ($Z_{\mathrm{bottom}}=0.5$ and $Z_{\mathrm{top}}=94.5$)}, as customary in LB simulations.\fi

%and $Z_{max}^{min} = Z_{\mathrm{top}}^{\mathrm{bottom}} \pm Z_{c}$ where $Z_{c} = 1.5$ in lattice units.This ensures that there is always at least one lattice node of fluid between the squirmer and a solid object.  

\begin{figure}
\includegraphics[width=\columnwidth]{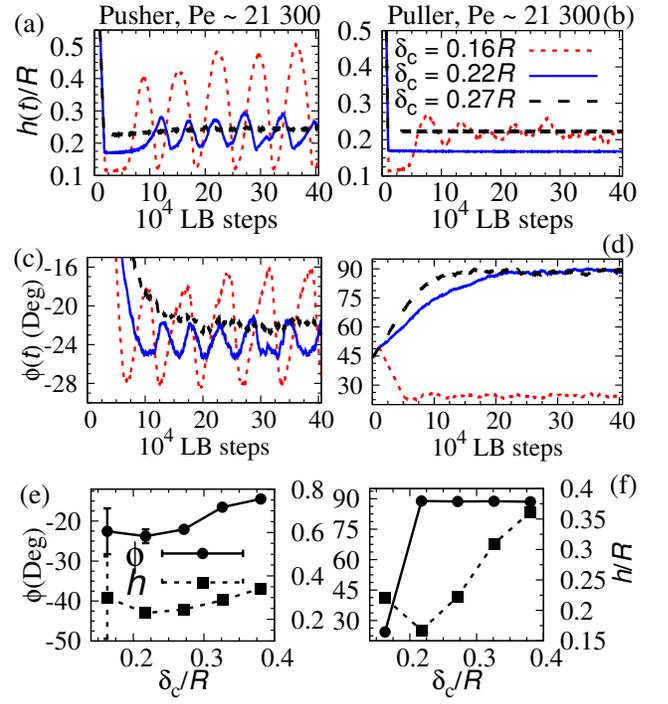}
\caption{Simulation results when the range of the soft repulsion is varied $\delta_c\approx 0.16R$ (dotted line), $\delta_c\approx 0.22R$ (solid line) and $\delta_c\approx 0.27R$ (dashed line), with $\mathrm{Pe}\approx 2\times 10^4$. Time development of $h(t)/R$ (a) for a pusher and (b) for a puller as well as $\phi(t)$ for (c) a pusher and (d) a puller. (The sudden change in behaviour for a puller is discussed in the text). The steady state $h$ and $\phi$ as a function of $\delta_c$ for (e) a pusher (the errorbars gives the amplitude of the oscillations) and (f) a puller. (Initial conditions: $h_0\approx 1.1R$ and $\phi_0=45.0^{\circ}$; $\mathrm{Pe}\approx 2\times 10^4$).}
\label{Fig:GapAngle1}
\end{figure}

{\it Results:} In Fig.~\ref{Fig:Cartoon}(c,d) we plot the evolution of the dimensionless gap size $\epsilon=h(t)/R$ and of the angle $\phi$ between the squirmer direction and the surface plane (Fig.~\ref{Fig:Cartoon}(a)). %In steady state swimming, the puller (pusher) points towards (away) from the no-slip surface ($\phi (t)$ in Fig.~\ref{Fig:Cartoon}(c,d)). 
For a puller, after an initial collision with the soft repulsive wall, the hydrodynamically induced torques rotate the particle so that it settles to swim parallel to the no-slip wall (beyond the excluded volume interaction range; dot-dashed line in Fig.~\ref{Fig:Cartoon}(c)) with a distance $h\sim 0.2R$, in very good agreement with theoretical predictions~\cite{ishimoto13}. In steady state the puller points towards the surface, $\phi\sim 24$ degrees (Fig.~\ref{Fig:Cartoon}(c)).
For a $\beta=-5$ pusher, previous theories based only on hydrodynamic interactions predict no stable swimming near a surface~\cite{ishimoto13}. However, in experiments, phoretic swimmers which are thought to be pushers for mechanistic reasons~\cite{das15}, are typically observed to accumulate and undergo stable swimming at no-slip surfaces~\cite{brown14, brown15}. Strikingly, our simulations (Fig.~\ref{Fig:Cartoon}(d)), show a stable periodic orbit both in $h(t)$ and $\phi(t)$. During a collision with the soft-repulsive wall, hydrodynamic torques reorient the pusher (Fig.~\ref{Fig:Cartoon}(d)) leading to it swimming away from the wall ($\phi$  is on average  $<0$ in Fig.~\ref{Fig:Cartoon}(d)). This much is expected; surprisingly, the long range hydrodynamic interactions between the swimmer and the wall, lead to another reorientation of the pusher so that it starts to swim towards the wall again. The cycle repeats leading to the the hydrodynamic oscillations near the no-slip surface (Fig.~\ref{Fig:Cartoon}(d)). While $\phi(t)<0$ during the oscillations, significant part of the trajectory $h(t)$ is spent beyond the external repulsion range (dot-dashed line in Fig.~\ref{Fig:Cartoon}(d)). Experimentally, these oscillations would be difficult to distinguish from true, steady-state trapping, providing a potential explanation for the experimental observations.
%(Note that a cyclic swimming and decaying cyclic swimming in $h$ has been reported for {\it pullers} in refs.~\cite{li14} and~\cite{ignacio1}.)
Interestingly, for both pusher and puller dynamics, the hydrodynamically induced attraction is strong enough to resist the effects of thermal noise (see Fig.~\ref{Fig:GapAngle1}(a,b)). Decreasing $\beta$ reduces the strength of hydrodynamic torques~\cite{brown15}, we observed no trapping for $\beta = 0,~\pm 2$ (see SI, Fig. S1).

The external soft repulsion, and in particular its range, plays a key role in determining the swimming dynamics. This can be seen from the $\phi(t)$ and $h(t)$ curves presented in Fig.~\ref{Fig:GapAngle1}(a-d), for different repulsive ranges ($\delta_c=0.16R,~0.22R$ and $0.27R$): these simulations include the effect of thermal noise, with $\mathrm{Pe}\approx 2\times 10^4$, and were all initialised with $h_0\approx 1.1R$ and $\phi_0 = 45^{\circ}$. For all ranges considered, both the pusher and the puller are found to swim near the surface (Fig.~\ref{Fig:GapAngle1}(a,b)). However, the hydrodynamic oscillations in the pusher dynamics (visible both in $h(t)$ and $\phi(t)$~Fig.~\ref{Fig:GapAngle1}(a,c)) are suppressed when the repulsive range is increased, and disappear altogether for $\delta_c=0.27R$ (Fig.~\ref{Fig:GapAngle1}). In this case, for both pusher and puller steady state $h<\delta_c$.
%(i.e. they lie for a significant time of their trajectoris above the dashed line in Fig.~\ref{Fig:GapAngle1} which indicates the repulsive range):
%with $h$ increasing with $\delta_c$, (Fig.~\ref{Fig:GapAngle1}(e,f)). 
The plot of the swimming orientation $\phi(t)$ (Fig.~\ref{Fig:GapAngle1}(c,d)) confirms the absence of oscillations: the pusher swims by keeping a stable orientation tilted away from the wall, with $\phi$ slightly decreasing when $\delta_c$ is increased (Fig.~\ref{Fig:GapAngle1}(e)); the puller instead is rotated by hydrodynamic torques to point towards the wall, so that $\phi\sim 90^{\circ}$ (this is always the case as soon as $\delta_c \ge 0.22R$ (Fig.~\ref{Fig:GapAngle1}(f))).

\begin{figure}
\includegraphics[width=\columnwidth]{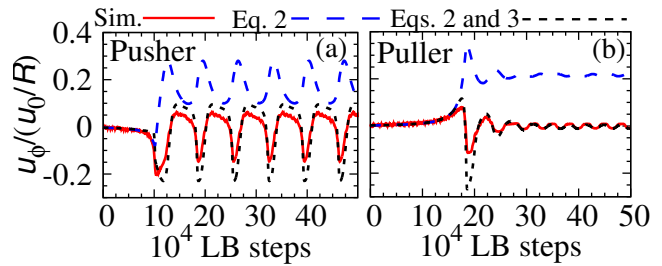}
\caption{Observed simulation (solid line), far field (dashed line) and combined near and far-field (dotted line), results for $u_{\phi}(t)/(u_0/R)$ for (a) pusher and (b) puller. The repulsive range  was $\delta_c\approx 0.16R$ (initial conditions $h_0\approx 3.25R$ and $\phi_0 = 10^{\circ}$; no thermal noise).}
\label{Fig:Vel1}
\end{figure}

Most existing theories of swimmer hydrodynamics rely on the far-field approximation which is based on the velocity field a swimmer generates at distances which are large with respect to its size. The far-field approximation can be adapted to include a no-slip wall~\cite{spagnolie12}: as a result one obtains the following expressions for the time derivative of $\phi$ at a given time,
\begin{eqnarray}
  \frac{\mathrm{d}\phi}{\mathrm{d}t}&=&\frac{u_0}{R}\left[\frac{9\beta\sin(2\phi)}{64}\left(\frac{R}{h^{\prime}}\right)^3
    -\frac{3\cos\phi}{16}\left(\frac{R}{h^{\prime}}\right)^4\right]\label{eq:ff} %\nonumber \\
  %\frac{\mathrm{d}x}{\mathrm{d}t}&=&u_0\left[\cos\phi\frac{9\beta\sin(2\phi)}{32}\left(\frac{R}{h^{\prime}}\right)^2
  %  - \frac{\cos\phi}{8}\left(\frac{R}{h^{\prime}}\right)^3\right]\label{eq:ff} 
 % \frac{dz}{dt}&=&u_0\left[\sin\phi-\frac{9\beta\left(1-3\sin^2\phi\right)}{32}\left(\frac{R}{h^{\prime}}\right)^2
 %   - \frac{\sin\phi}{2}\left(\frac{R}{h^{\prime}}\right)^3\right]\nonumber ,
\end{eqnarray}
where $\beta = \tfrac{B_2}{B_1}$ and $h^{\prime}$ is the distance from the centre of the particle to the confining wall ($h^{\prime} = h+R$ in Fig.~\ref{Fig:Cartoon}(a)). Alternatively, one may use lubrication theory to compute the following prediction, based on near-field hydrodynamics~\cite{ZottlThesis, cichocki98},
  \begin{eqnarray}
    \frac{\mathrm{d}\phi}{\mathrm{d}t} &=& -\frac{3u_0}{2R}\left(1+\beta\sin\phi\right)\cos\phi + {\cal{O}}\left(\frac{1}{\log\epsilon^{-1}}\right)\label{eq:nf} %\nonumber\\
%    \frac{\mathrm{d}x}{\mathrm{d}t} &=& -\frac{7u_0}{10}\left(1+\beta\sin\phi\right)\cos\phi + {\cal{O}}\left(\frac{1}{\log\epsilon^{-1}}\right)\label{eq:nf},
  \end{eqnarray}
  where $\epsilon = (h^{'}-R)/R$~\cite{Suppl}. %[See SI for far-field equations for $\tfrac{\mathrm{d}x}{\mathrm{d}t}$ and $\tfrac{\mathrm{d}z}{\mathrm{d}t}$, and for details of the derivation the near-field equations  and of how the far and near-field contributions can be combined.]}

To allow detailed comparison between our model and the theoretical predictions (Eqs. 2 and 3), we carried out simulations starting with $h_0\approx 3.25R$ and $\phi_0=10^{\circ}$ and at each point we calculated the expression for $d\phi/dt$ %, $dx/dt$ and $dz/dt$
either directly from our numerics, or by substituting the instantaneous values of $h(t)$ and $\phi(t)$ into Eqs.~\ref{eq:ff} and~\ref{eq:nf}: these two equations respectively provide the far- and near-field estimate of the system evolution given its current state and can be combined by means of a matched asymptotic expansion~\cite{Suppl}.

%First, we discuss the results for the angular velocity, $d\phi/dt$ (Fig.~\ref{Fig:Vel1}).
 For early times, there is good agreement between the rotational dynamics, $d\phi/dt$, predicted by the far-field approximation and that found in our direct numerical simulations (Fig.~\ref{Fig:Vel1}): we observe a decrease (pusher) and increase (puller) of $\phi(t)$ from the initial $\phi_0=10^{\circ}$. Later on, the far-field estimate no longer captures the dynamics observed in simulation. In steady state, the far field predicts $\tfrac{d\phi}{dt} > 0$ while simulations show no net motion of $\phi(t)$, as shown in Fig.~\ref{Fig:GapAngle1}(c,d). When incorporating the near-field contribution,  we observe very good agreement between the theory and our simulations, including the trapping of the puller and the oscillations of the pusher (Fig.~\ref{Fig:Vel1}). This result can be understood by noting that the far- and near-field contributions are qualitatively different. In the far-field, a pusher swims stably parallel to the wall, whereas a puller rotates until it is perpendicular to it~\cite{spagnolie12}. In the near-field, it is the puller which swims stably along the wall, pointing slightly towards it~\cite{ishimoto13}, whereas the pusher has no stable swimming solution. In our simulations, the particle is trapped close to the wall, so near-field hydrodynamics dominates, although far-field contributions are non-negligible.~\cite{Suppl}
  An analysis of the dynamics of approach to the surface, $dz/dt$~\cite{Suppl}, leads to similar conclusions: prior to interacting with the repulsive wall, the far-field works well; when the repulsive interaction is reached, there is a notable disagreement (see SI, Fig. S2).

\begin{figure}
\includegraphics[width=\columnwidth]{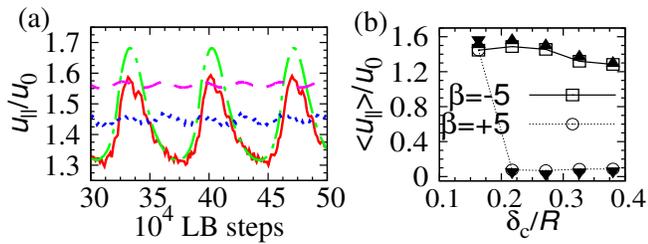}
\caption{(a) Simulations and far field results for the observed steady state swimming speed along the wall $u_{||}(t)/u_0$ for a pusher ($\beta = -5$); simulations (solid line) and far field predictions (dot-dashed line) and puller ($\beta= +5$); simulations (dotted line) and far field prediction (dashed line). ($\delta_c \approx 0.16R$; initial conditions: $h_0\approx 1.1R$, $\phi_0 = 45^{\circ}$; no thermal noise). (b) Time averaged $\langle u_{||}\rangle/u_0$ as a function of the repulsion range $\delta_c$, from simulations (open symbols) and far field calculations (closed symbols), for both pushers (squares and upward triangles) and pullers (circles and downward triangles). Initial conditions: $h_0\approx 1.1R$, $\phi_0=45^{\circ}$; $\mathrm{Pe}\approx 2\times 10^4$.}
\label{Fig:Vel2}
\end{figure}

For movement along the wall ($dx/dt\equiv u_{\parallel}$), the simulations and far field predictions agree reasonably well at all times %but can be improved by including near-field effects
(Fig.~\ref{Fig:Vel2}(a); similar conclusions were reached by Spagnolie and Lauga who studied the dynamics of a swimmer before collisions with the wall~\cite{spagnolie12}). 
The steady state velocities for both swimmers are considerably larger than in the bulk, i.e. the presence of a surface accelerates the motion ($\sim 50$\% increase, see Fig.~\ref{Fig:Vel2}(a)). %{\color{red} In the near-field of $u_{\parallel}$ the order unity term vanishes, and the leading order decays as $\mathcal{O}(1/\log\epsilon)$ (see SI for details). In Fig.~\ref{Fig:Vel1} $\epsilon \sim 0.1$, thus the near field effects could be expected to contribute to slowing down the swimmer.}
The increase in the swimming speed near a solid surface  can be understood intuitively by considering the swimming mechanism. The pusher is propelled from behind, thus when pointing away from the surface it is pushing the fluid flow against a solid wall -- this should enhance the swim speed, as predicted for swimmers in porous media~\cite{ledesma13, brown15}. The speed increase is retained for the periodic swimming. %Now $u(t)$ oscillates, as could be expected, but retains, on average, $u(t)\sim 1.5u_0$ (solid line in Fig.~\ref{Fig:Vel1}(e)). 
The speedup of the puller can be understood in a similar way: the squirmer is now oriented towards the wall so by pulling inward along its swimming axis, it pulls itself along the wall. %Although our speed increase occurs close to a solid wall, it is interesting to note that
Recent experiments reported an enhanced swimming speed up to $\sim 2u_0$ for Janus colloids on a water-air interface~\cite{wang15}. 

Increasing the range of the repulsive interaction leaves the pusher dynamics along the surface mostly unaffected: we find that $u_x>u_0$ for all the interaction ranges considered, as shown in Fig.~\ref{Fig:Vel2}(a,b).
The case of the puller is very different, as any repulsive interaction extending past the equilibrium swimming distance $\sim 0.2R$ leads to hydrodynamic torques orienting the particle towards the wall (Fig.~\ref{Fig:GapAngle1}(d)); see also SI, Fig. S3, which shows that the far-field approximation for the tangential speed remains good in this case as well). The reorientation occurs independently of initial conditions (in Fig.~\ref{Fig:Vel2}(b) the initial angle is almost tangential, $\phi_0=10^{\circ}$), and leads to a dramatic slowing down of the particle, whose motion virtually comes to a standstill when $\delta_c \ge 0.22R$ (see Fig.~\ref{Fig:Vel2}(b)). %For the movement along the confining surface the far field predictions agree remarkably well with what is observed from simulations (Fig.~\ref{Fig:Vel2}(b)) for all repulsion ranges considered.

In all the cases we have considered, the rotational motion has a fundamental role in the dynamics of the particle, and this is affected by $\delta_c$. The soft repulsion only slows down the particle movement along the surface normal (as visible from SI, Fig. S2 for $dz/dt$), and it {\it does not} create any torques; therefore any rotational motion of the particle only arises from the combination of hydrodynamic and Brownian forces. 

{\it Conclusions:}
We have presented a study of fully resolved spherical squirmers swimming between two solid walls, using a microscopic model which prescribes a slip velocity at the particle surface. Our results show that repulsive interactions, which have been neglected in previous theories of swimmers interacting with surfaces, play a very important role in the squirmer's dynamics. First, they can stabilise hydrodynamic oscillations of a pusher close to the wall.
A recent systematic investigation has demonstrated that in the parameter range we consider ($\beta \sim -5$ or below) there is no stable bound state with the pusher swimming near the wall~\cite{ishimoto13}.
  While experiments routinely observe that bacteria (which are known to be pushers) or phoretic swimmers (which are thought to be pushers) are attracted to and swim near flat surfaces~\cite{berke08,brown14}. One way to reconcile these results is if the trajectory of the swimmer at late times is oscillatory (a limit cycle in the $(h,\phi)$ plane) instead of having constant velocity (a stationary point in the $(h,\phi)$ plane). While this conclusion should hold qualitatively for several different pusher swimmers, we note that a spherical squirmer model does not provide a quantitatively accurate description of a rod-like bacterial swimmer such as {\it E.coli}, so that the details of its hydrodynamic oscillations may in practice differ from those presented here.

Second, we find that the swim velocity of a pusher is much increased with respect to the bulk limit: this behaviour can be understood as the particle, on average, is directed away from the wall and pushes on it, enhancing its speed. Third, we find that the tangential velocity of a puller slows down dramatically with the range of the repulsive interaction with the wall. 
Our results critically require near-field hydrodynamics, as the far-field approximation poorly captures the rotational dynamics we observe. 

Our findings further imply that the existence and extent of steric or electrostatic repulsion of the wall could be tuned to control properties such as the number density and speed of active particles near a surface. Experimentally this could be achieved, by varying either the buffer concentration (for electrostatic repulsion) or the polymer coverage of the surface (for steric repulsion). These predictions should be testable with experiments using bacterial swimmers or artificial microswimmers, although for phoretic particles one may need to first estimate the effect of chemical gradients, here neglected, on the dynamics~\cite{uspal15, brown15, theurkauff12, bickel13, ginot15}. 

{\it Acknowledgements:} We thank Andreas Z{\"o}ttl and Joost de Graaf for fruitful discussions. This work was funded by EU intra-European fellowship 623637 DyCoCoS FP7-PEOPLE-2013-IEF and UK EPSRC grant EP/J007404/1.

%\begin{thebibliography}{99}

%\end{thebibliography}

%\bibliography{REFERENCELIST}

%

\onecolumngrid
\newpage

\renewcommand{\thefigure}{S\arabic{figure}}

\setcounter{equation}{0}
\setcounter{figure}{0}

\newtoks\rowvectoks
\newcommand{\rowvec}[2]{%
 \rowvectoks={#2}\count255=#1\relax
 \advance\count255 by -1
 \rowvecnexta}
\newcommand{\rowvecnexta}{%
 \ifnum\count255>0
 \expandafter\rowvecnextb
 \else
 \begin{pmatrix}\the\rowvectoks\end{pmatrix}
 \fi}
\newcommand\rowvecnextb[1]{%
 \rowvectoks=\expandafter{\the\rowvectoks&#1}%
 \advance\count255 by -1
 \rowvecnexta}

%\maketitle
\begin{center}
  {\Large \bf  Supplementary material for hydrodynamic oscillations and variable swimming speed in squirmers close to repulsive walls}
\end{center}
\medskip

\section{Soft repulsive potential at the wall}
We use the following potential between the particle and the wall,
\begin{eqnarray}
 V_{cs}(h) = V(h) - V(h_{c}) - (h-h_c)\left.\frac{\partial V(h)}{\partial h}\right|_{h=h_c},
\end{eqnarray}\label{repulsion}
where the wall-particle-surface separation $h = Z_{max}^{min}  \pm z - R$, and
\begin{equation}
V(h) = \epsilon\left(\sigma/h\right)^\nu.
\end{equation}
$V_{cs}$ has been cut-and-shifted to ensure that the potential and force go smoothly to zero at $h=h_c$. Parameters were chosen as $\epsilon = 0.004$, $\sigma = 0.1$, $\nu=1.0$. By choosing $Z_{\mathrm{max}} = Z_{\mathrm{top}} - (\delta_c - h_c)$ ($Z_{\mathrm{min}} = Z_{\mathrm{bottom}} + (\delta_c - h_c)$) and keeping $h_c=0.5$SU constant, we can have a well defined repulsion range $\delta_c$, while keeping $h_c$ and thus the potential form constant (Fig.1(b) in the main text). For the calculation of the gap size between the squirmer and the solid surface $h$ (Fig. 1(a) in the main text), we define the wall location half-way between the  solid node and first fluid node ($Z_{\mathrm{bottom}}=0.5$ and $Z_{\mathrm{top}}=94.5$), as customary in LB simulations.  

\section{Far-field approximation}

The far-field approximation is based on the velocity field which a swimmer generates at distances which are large with respect to its size. The far-field approximation can be adapted to include a no-slip wall~\cite{spagnolie12}: as a result one obtains the following expressions for the time derivative of the positions parallel $x$ and perpendicular $z$ as,
\begin{eqnarray}
  %\frac{\mathrm{d}\phi}{\mathrm{d}t}&=&\frac{u_0}{R}\left[\frac{9\beta\sin(2\phi)}{64}\left(\frac{R}{h^{\prime}}\right)^3
%    -\frac{3\cos\phi}{16}\left(\frac{R}{h^{\prime}}\right)^4\right]\label{eq:ff} %\nonumber \\
  \frac{\mathrm{d}x}{\mathrm{d}t}&=&u_0\left[\cos\phi\frac{9\beta\sin(2\phi)}{32}\left(\frac{R}{h^{\prime}}\right)^2
    - \frac{\cos\phi}{8}\left(\frac{R}{h^{\prime}}\right)^3\right]\\ 
  \frac{dz}{dt}&=&u_0\left[\sin\phi-\frac{9\beta\left(1-3\sin^2\phi\right)}{32}\left(\frac{R}{h^{\prime}}\right)^2
    - \frac{\sin\phi}{2}\left(\frac{R}{h^{\prime}}\right)^3\right]\nonumber ,
\end{eqnarray}
where $\beta = \tfrac{B_2}{B_1}$ and $h^{\prime}$ is the distance from the centre of the particle to the confining wall ($h^{\prime} = h+R$ in Fig. 1(a) in the main text).

\begin{figure}
  \includegraphics[width=0.4\columnwidth, angle=270]{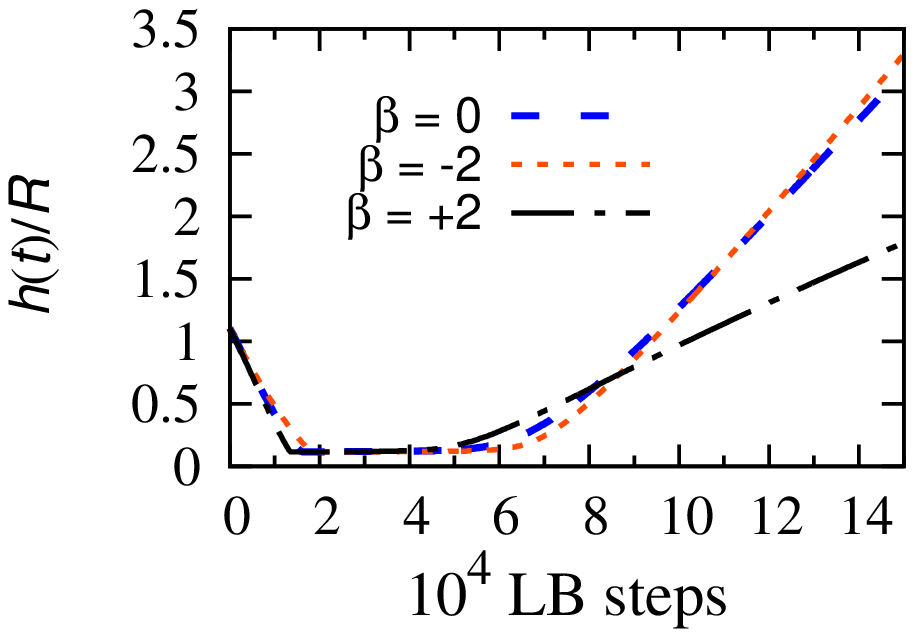}
  \caption{Observed simulation results for $h(t)/R$ for a neutral squirmer ($\beta = 0$), for a pusher ($\beta = -2$) and for a puller ($\beta = +2$). The repulsive range was $\delta_c \approx 0.16R$ and $\mathrm{Pe}\approx 2\times 10^4$.} 
  \label{Fig:S3}
\end{figure}

\begin{figure}
\includegraphics[width=\columnwidth]{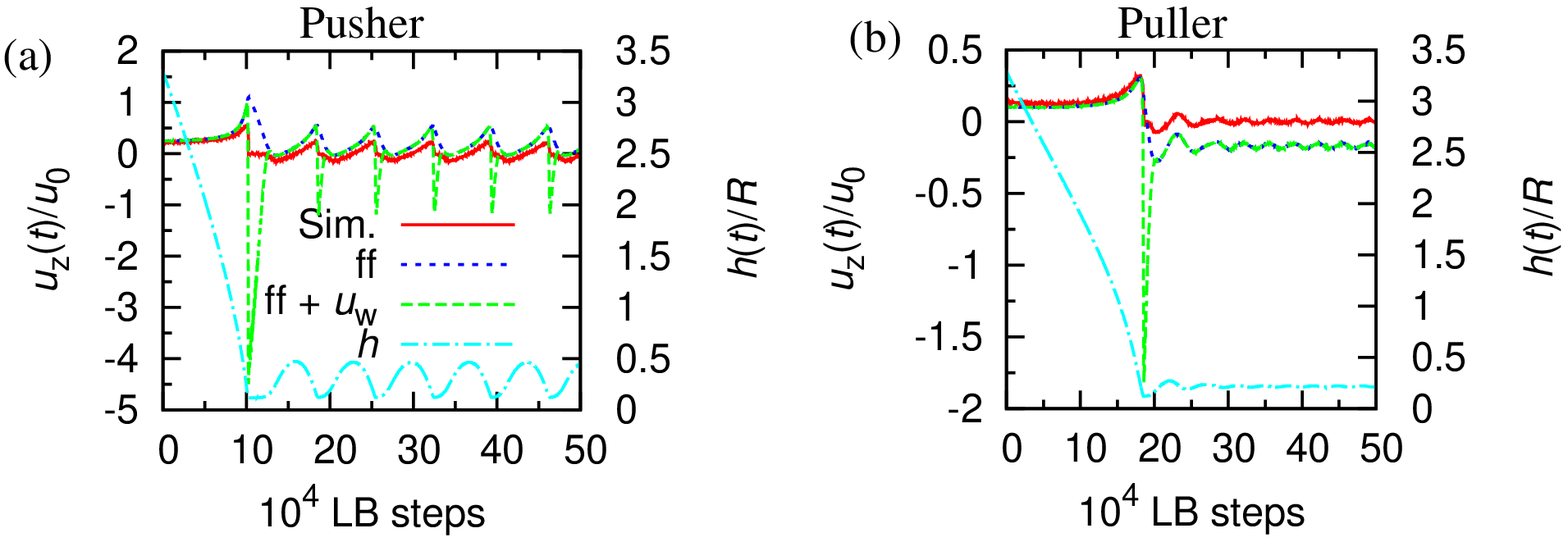}
\caption{Observed simulation results (solid line and dot-dashed line) and
far field results (dotted line) for $u_z(t)/u_0$ for (a) pusher and (b)
puller. The disagreement between simulations and far-field
approximation persists after correcting the latter to include the
repulsive interaction from the wall, via an extra normal velocity
equal to $u_w=-\tfrac{1}{\gamma}\tfrac{\partial V(h)}{\partial h}$ (where
$\gamma = 6\pi\eta R$). The repulsive range  was $\delta_c\approx
0.16R$ (initial conditions $h_0\approx 3.25R$ and $\phi_0 = 10^{\circ}$;
no thermal noise).}
%Observed simulations results and far field results for $u_z(t)/u_0$ for (a) pusher and (b) puller. The repulsion range was $\delta_c\approx 0.16R$ (initial conditions $h_0\approx 3.25R$ and $\phi_0 = 10^{\circ}$; no thermal noise)}
\label{Fig:S1}
\end{figure}

\begin{figure}
\includegraphics[width=0.5\columnwidth]{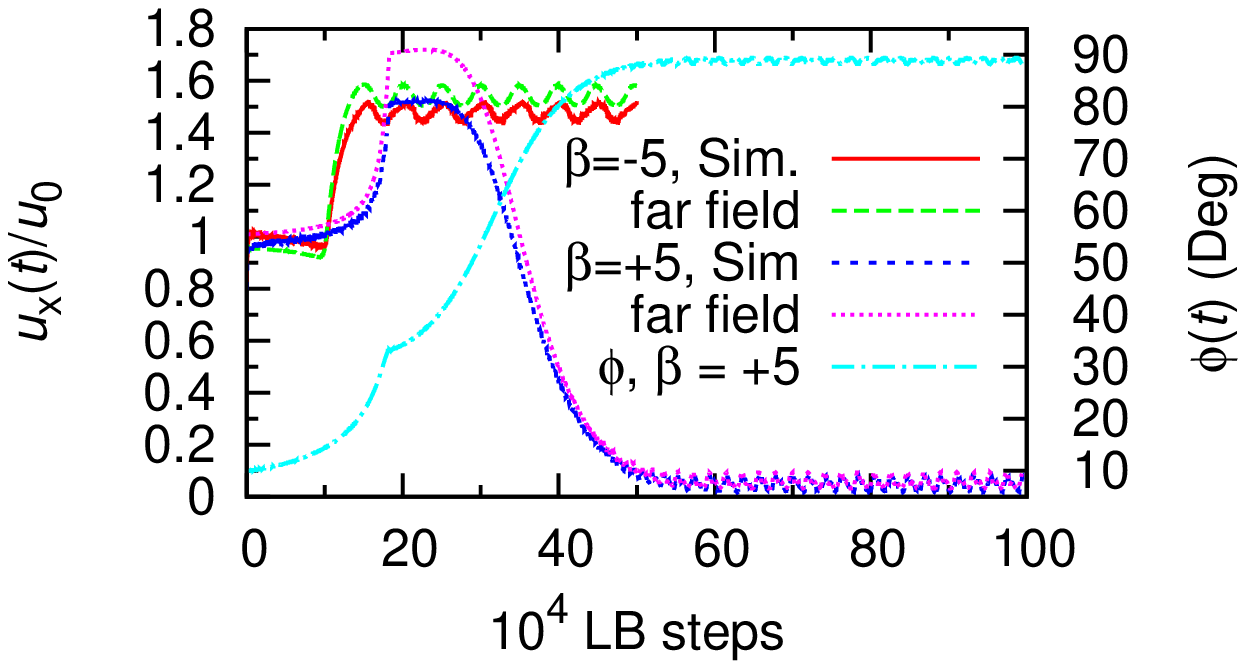}
\caption{Simulations and far field results for the velocity along the wall $u_{x}(t)/u_0$ for pusher ($\beta = -5$) and puller ($\beta= +5$) as well as $\phi(t)$ for a puller ($\beta=+5$), when the repulsion range was $\delta_c\approx 0.22R$ (initial conditions $h_0\approx 3.25R$, $\phi_0=10^{\circ}$; no thermal noise).}
\label{Fig:S2}
\end{figure}

\cleardoublepage

\section{Lubrication Results}

\begin{figure}[]
\centering
\includegraphics[width=8.5 cm]{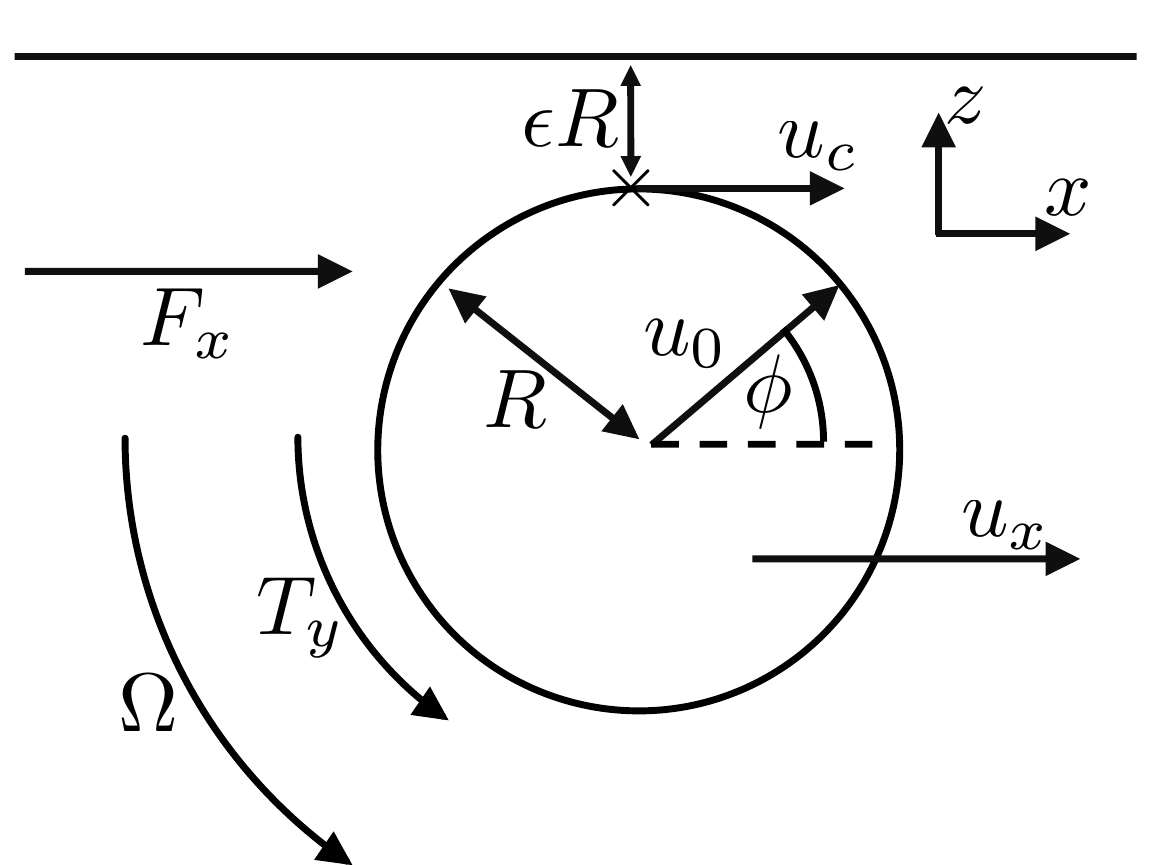}
\caption{Definition of the geometry of a spherical squirmer of radius $R$ next to a plane surface. The $\hat{\mathbf{y}}$ direction is into the page. In the bulk, the squirmer would move at speed $u_0$ along the direction shown, at angle $\phi$ from the horizontal. The squirmer is axisymmetric around this axis. $u_c$ is the slip velocity at the point of closest approach, indicated by $\times$. At this point, the squirmer wall gap is $\epsilon R$. The fluid flow generates a force $F_x$ and torque $T_y$ on the squirmer, around its centre, which generate translation at speed $u_x$ along the x-axis, and rotation, at angular velocity $\Omega$ around the y-axis.}
\label{fig:geometry}
\end{figure}

In Ref.~\cite{ishikawa06}, the wall-parallel force $F_x$ and torque $T_y$ on a squirmer moving near a no-slip boundary is calculated. We repeat these calculations in Section~\ref{T_calc}. We obtain different prefactors for $F_x$ and $T_y$ compared to Ref.~\cite{ishikawa06}, but we agree as to the functional form. In summary, our corrected results for a squirmer oriented at angle $\phi$ away from the parallel to the plane (see Fig.~\ref{fig:geometry}) are
\begin{align}
	F_x\,=\,-\frac{4\pi\eta Ru_c}{5}\log(1/\epsilon)+\mathcal{O}(1)\,, \label{rotating squirmer force}
\end{align}
\begin{align}
	T_y\,=\,\frac{16\pi\eta R^2u_c}{5}\log(1/\epsilon)+\mathcal{O}(1)\,. \label{rotating squirmer torque}
\end{align}
Here, $u_c$ is the surface slip velocity (parallel to the wall) at the point of closest approach ($\times$) between the squirmer and the wall, which, for the squirmer defined in Eq. 1 of the main text is
\begin{align}
	u_c=-u(\pi/2-\phi)=-\frac{3}{2}u_0\cos{\phi}\left(1+\beta\sin\phi\right)\,. \label{u_c}
\end{align}
From standard results for the drag on a sphere near a wall~\cite{kim13} the force and torque give simple expressions for the {\it total} rotation $\Omega$ and speed $u_x$ of the squirmer
\begin{align}
	\Omega=\frac{\mathrm{d}\phi}{\mathrm{d}t}=\frac{u_c}{R}+\mathcal{O}(1/\log\epsilon)\,, \label{Omega}
\end{align}
\begin{align}
	u_x=\frac{\mathrm{d}x}{\mathrm{d}t}=0+\mathcal{O}(1/\log\epsilon)\,. \label{u_x}
\end{align}
or, in terms of $u_0$ and $\phi$
\begin{eqnarray}
    \frac{\mathrm{d}\phi}{\mathrm{d}t} &=& -\frac{3}{2}\frac{u_0}{R}\cos{\phi}\left(1+\beta\sin\phi\right) + {\cal{O}}\left(\frac{1}{\log\epsilon}\right)\nonumber\\
    \frac{\mathrm{d}x}{\mathrm{d}t} &=& 0+{\cal{O}}\left(\frac{1}{\log\epsilon}\right)\label{eq:nf}.
  \end{eqnarray}
In other words, the term of order unity in the total translational motion of a squirmer near a wall vanishes, and the leading order speed decays as $\mathcal{O}(1/\log\epsilon)$ as the squirmer approaches the wall. It is not possible to calculate the numerical value of this term from lubrication theory, since it depends on longer-range interactions between the whole squirmer and the wall~\cite{kim13}. Since this logarithmic decay is very weak, the leading order term will remain comparable to $u_0$ except for squirmers extremely close to surfaces. In the current simulations $\epsilon\gtrsim0.1$, giving $|1/\log\epsilon|\gtrsim 0.4$, which is not small. Hence, it is not contradictory that, in simulations, we see $u_x$ increase as the squirmer approaches the wall: this is probably because the swimmer does not approach the wall very closely in the simulations. The lubrication calculations merely predict that, for a sufficiently close approach, the translational speed of the squirmer will begin to decrease and eventually slow to zero. For the vertical speed, $u_z$, the term of order unity also vanishes~\cite{ishikawa06}, so we do not calculate $u_z$ here. For the rotational motion, the next-to-leading order term also decays as $\mathcal{O}(1/\log\epsilon)$, so it will also be significant. 

We can provide an intuitive justification for Eq.~\eqref{Omega}-\eqref{u_x}. As the swimmer gets closer and closer to the surface, most of the viscous dissipation will occur in the thin region around the contact point. We would therefore expect the solution to minimise the dissipation in this region. This can be done by ensuring that there is no difference in fluid velocity between the particle and the plane surface at this point of contact. Hence, the total velocity on the particle surface, taking into account the slip velocity and the solid-body motion of the particle should, in the limit of infinitessimal gap size, approach zero, i.e.,
\begin{align}
	\lim_{\epsilon\rightarrow 0}\left(u_c+u_x-R\Omega\right)=0\,. \label{condition}
\end{align}
This condition is satisfied (but not uniquely) by Eq.~\eqref{Omega}-\eqref{u_x}. It is not satisfied by the original result derived in Ref.~\cite{ishikawa06}.

\section{Calculation of Lubrication Force and Torque \label{T_calc}}

We briefly repeat here the lubrication calculations of Ref.~\cite{ishikawa06}, to obtain the results in Eq.~\eqref{rotating squirmer force}-\eqref{rotating squirmer torque}. This calculation is identical to the standard calculation of the forces and torques of a no-slip sphere near a surface~\cite{kim13}, except for the new boundary condition on the sphere surface introduced by the finite slip velocity. We first define a cylindrical coordinate system $(\rho^{*}, z, \psi )$, with $\rho^{*2}=x^2+y^2$ and $\tan{\psi}=y/x$, where the origin of the coordinate system is the point on the plane immediately above the squirmer's centre. The boundary of the squirmer is defined by $z=h(\rho^*,\phi)$, and in the vicinity of the contact point is given by
\begin{align}
h\,=\,-R\left(\epsilon+\frac{\rho^{*2}}{2R^2}+\mathcal{O}\left(\frac{\rho^{*4}}{R^2}\right)\right)\,.
\end{align}
To ensure that the equations of motion are all of order unity, we use the dimensionless stretched variables $X,~Y,~Z,~H,~\rho$, with the scaling
\begin{align}
\epsilon^{1/2}RX=x,\,\,\,\,\epsilon^{1/2}RY=y,\,\,\,\, \epsilon^{1/2}&R\rho=\rho^*\\ 
\epsilon RZ=z\,, \,\,\,\,\,\,\,\,\,\epsilon RH=h&\,.
\end{align}
The stretched height $H$ is
\begin{align}
H=-1-\frac{\rho^2}{2}+\mathcal{O}(\epsilon)\,.
\end{align}
The fluid velocity field $\mathbf{u}$ has $x,~y,~z$ components $u$, $v$ and $w$ respectively. In the stretched coordinate system, the Stokes equations are
\begin{align}
&\eta\left(\epsilon\nabla^2_\parallel +\frac{\partial^2}{\partial Z^2}\right)\mathbf{u}=R\rowvec{3}{\epsilon^{1/2}\dfrac{\partial p}{\partial X},}{\epsilon^{1/2}\dfrac{\partial p}{\partial Y},}{\dfrac{\partial p}{\partial Z}}\,,\label{Stokes 1}\\
&\epsilon^{1/2}\left(\frac{\partial u}{\partial X}+\frac{\partial u}{\partial Y}\right)+\frac{\partial w}{\partial Z}=0\,. \label{Stokes 2}
\end{align}
where $\nabla_\parallel=\hat{\mathbf{x}}\partial/\partial X + \hat{\mathbf{y}}\partial/\partial Y$ and with fluid viscosity $\eta$ and pressure $p$. No-slip boundary conditions apply on the plane surface: $\{u,v, w\}|_{Z=0}=0$, and we write the boundary velocity on the upper surface as $u(Z=H)=U$, $v(Z=H)=V$ and $w(Z=H)=W$. Expanding the boundary conditions as power series in orders of $\epsilon^{1/2}$ around $\rho=0$, we have
\begin{align}
 U=&u_c+\mathcal{O}(\epsilon^{1/2})\,,\\
 V=&0+\mathcal{O}(\epsilon^{1/2})\,,\\
W=&0-u_c\epsilon^{1/2}\rho\cos{\psi}+\mathcal{O}(\epsilon)\,.
\end{align}
Performing a similar expansion for the bulk velocity and pressure gives
\begin{align}
u&=u_0+\epsilon^{1/2}u_1+\mathcal{O}(\epsilon)\,,\\
v&=v_0+\epsilon^{1/2}v_1+\mathcal{O}(\epsilon)\,,\\
w&=0+\epsilon^{1/2}w_0+\epsilon w_1+\mathcal{O}(\epsilon^{3/2})\,,\\
p=\epsilon^{-3/2}&\left[ p_0+\epsilon^{1/2}p_1+\mathcal{O}(\epsilon)\right]\,.
\end{align}
From Eq.~\eqref{Stokes 1}, $p_0$ is independent of $Z$. Solving for $u_0$ and $v_0$ in Eq.~\eqref{Stokes 1}-\eqref{Stokes 2} then gives
\begin{align}
u_0&\,=\,\frac{R}{2\eta}\frac{\partial p_0}{\partial X}(Z-H)Z+\frac{Z}{H}u_c\,,\\
v_0&\,=\,\frac{R}{2\eta}\frac{\partial p_0}{\partial Y}(Z-H)Z\,.
\end{align}
Combining these solutions and using the equation of continuity (Eq.~\eqref{Stokes 2}) yields, after some algebra
\begin{align}
\frac{H^3R}{12\eta}\nabla^2_\parallel p_0 - \frac{H^2R}{4\eta}\mathbf{\uprho}\cdot\nabla_\parallel p_0-\frac{1}{2}u_c\rho\cos\psi=0\,.
\end{align}
Inserting the ansatz $p_0=q_0(\rho)\cos{\psi}$ then gives an equation in terms of $\rho$ alone
\begin{align}
\frac{H^3R}{12\rho^2\eta}\frac{\partial}{\partial \rho}\left(\rho\frac{\partial q_0}{\partial \rho}\right) - \frac{H^3R}{12\rho^3\eta}q_0-\frac{H^2R}{4\eta}\frac{\partial q_0}{\partial \rho}+\frac{1}{2}u_c=0\,.
\end{align}
which has the particular solution
\begin{align}
q_0=-\frac{6\eta\rho}{5 H^2R}u_c\,.
\end{align}
As discussed in~\cite{ishikawa06}, the conditions that $p_0$ be finite everywhere means that this is the only physically relevant solution. 

Next, we rewrite the $x, y$ velocities in the cylindrical polar coordinate system, i.e.,
\begin{align}
\mathbf{u}\,=\,u_\rho\hat{\mathbf{\uprho}}+u_\psi\hat{\mathbf{\uppsi}}+u_z\hat{\mathbf{z}}\,.
\end{align}
Using the ansatz 
\begin{align}
u_\rho&\,=\,\cos\psi \tilde{u}_\rho(\rho)\,,\\
u_\psi&\,=\,\sin\psi\tilde{u}_\psi(\rho)\,,
\end{align}
we obtain for the in-plane components
\begin{align}
\tilde{u}_\rho&\,=\,\frac{R}{2\eta}(Z-H)Zq_0'+\frac{Z}{H}u_c\,,\\
\tilde{u}_\psi&\,=\,-\frac{R}{2\eta}(Z-H)Z\frac{q_0}{\rho}-\frac{Z}{H}u_c\,,
\end{align}
where the prime indicates the radial derivative $\partial/\partial\rho$.

To obtain the total horizontal force $F_x$ on the swimmer, we integrate small elements of force over the  swimmer surface $S$, i.e.,
\begin{align}
F_x=\int_S {\rm d}F_x\,,
\end{align}
where~\cite{kim13}
\begin{align}
{\rm d}F_x=\hat{\mathbf{x}}\cdot\mathbf{\upsigma}\cdot\hat{\mathbf{n}}{\rm d}S\,,
\end{align}
with $\mathbf{\upsigma}$ the stress tensor, and ${\rm d}S$ an infinitesimal area element. We evaluate the stress tensor in cylindrical polar coordinates, but use spherical polar coordinates centred on the particle centre, with the polar angle $\chi=0$ at the point of closest approach, to specify the normal $\hat{\mathbf{n}}$. This gives
%\begin{widetext}
\begin{align}
{\rm d}F_x\,=\,\left[-p\sin\chi\cos\psi+\eta\left(\sin\chi\cos\psi\tau_{\rho*\rho*}-\sin\chi\sin\psi\tau_{\pi\rho*}+\cos\chi\cos\psi\tau_{\rho* z}-\cos\chi\sin\psi\tau_{\psi z}\right)\right]{\rm d}S\,.
\end{align}
where $\tau$ is the rate of strain tensor, with components (in the unstretched coordinates $z, \psi, \rho^*$)
\begin{align}
 \tau_{\rho*\rho*}&=2\frac{\partial v_{\rho}}{\partial \rho*}\,,\,\,\,\,\,\,\,\,\,\,\,\,\,\,\,\,\,\,\,\,\,\,\,\,\,\,\,\,\,\,\,\,
 \tau_{\psi \rho*}=\rho*\frac{\partial}{\partial \rho*}\left(\frac{v_\psi}{\rho*}\right)+\frac{1}{\rho*}\frac{\partial v_{\rho}}{\partial \psi}\,,\nonumber\\
 \tau_{\rho* z}&=\frac{\partial v_{\rho}}{\partial z}+\frac{\partial v_z}{\partial {\rho*}}\,,\,\,\,\,\,\,\,\,\,\,\,\,\,\,\,\,\,\,\,\,
\tau_{\psi z}=\frac{\partial v_\psi}{\partial z}+\frac{1}{\rho*}\frac{\partial v_z}{\partial \psi}\,,
\end{align}
and ${\rm d}S=R^2\sin\chi{\rm d}\chi{\rm d}\psi$ is the area increment in spherical polar coordinates. Inserting the expansions for the velocities and rescaling into the stretched coordinates gives
\begin{align}
{\rm d}F_x= &\left\{ -\epsilon^{-3/2}q_0\sin\chi\cos^2\psi+\frac{\eta}{R}\left[2\epsilon^{-1/2}\sin\chi\cos^2\psi\frac{\partial \tilde{u}_\rho}{\partial \rho} -\epsilon^{-1/2}\sin\chi\sin^2\psi\left(\rho\frac{\partial}{\partial \rho}\left(\frac{\tilde{u}_\psi}{\rho}\right)
-\frac{1}{\rho}\frac{\partial \tilde{u}_\rho}{\partial \psi}\right) + \right.\right.\\
& \left.\left. +\cos\chi\cos^2\psi\left(\epsilon^{-1}\frac{\partial \tilde{u}_\rho}{\partial Z}+\frac{\partial \tilde{u}_z}{\partial \rho}\right)-\cos\chi\sin^2\psi\left(\epsilon^{-1}\frac{\partial \tilde{u}_\psi}{\partial Z}-\frac{1}{\rho}\frac{\partial \tilde{u}_z}{\partial \psi}\right)\right]\right\}{\rm d}S\bigg|_{Z=H}\,.
\end{align}
Performing the integral over $\psi$ gives
\begin{align}
F_x=R^2\pi\int_0^{\pi}\left\{-\epsilon^{-3/2}q_0\sin\chi+\frac{\eta}{R}\left[2\epsilon^{-1/2}\sin\chi\frac{\partial \tilde{u}_\rho}{\partial \rho} -\epsilon^{-1/2}\sin\chi\left(\rho\frac{\partial}{\partial \rho}\left(\frac{\tilde{u}_\psi}{\rho}\right)
-\frac{1}{\rho}\frac{\partial \tilde{u}_\rho}{\partial \psi}\right) + \right. \right. \nonumber\\
\left. \left. +\cos\chi\left(\epsilon^{-1}\frac{\partial \tilde{u}_\rho}{\partial Z}+\frac{\partial \tilde{u}_z}{\partial \rho}\right)-\cos\chi\left(\epsilon^{-1}\frac{\partial \tilde{u}_\psi}{\partial Z}-\frac{1}{\rho}\frac{\partial \tilde{u}_z}{\partial \psi}\right)\right]\right\}\sin{\chi}{\rm d}\chi\bigg|_{Z=H}\,.
\end{align}
%\end{widetext}
To perform the integral over $\chi$, we expand to first order around $\chi=0$, giving $\sin\chi=\epsilon^{1/2}\rho+\mathcal{O}(\epsilon)$. The inner, lubrication region extends to some real distance $\rho_0^*$ of order the particle size, $\rho_0^*=DR$, where $D=\mathcal{O}(1)$ is an unknown constant which can be obtained by matching to the outer solution. In the stretched coordinate system, the corresponding limit is $\rho_0=D\epsilon^{-1/2}$. To lowest order 
\begin{align}
F_x=R^2\pi\int_0^{\rho_0}-q_0\rho^2+\frac{\eta\rho}{R}\left(\frac{\partial \tilde{u}_\rho}{\partial Z}\bigg|_{Z=H}-\frac{\partial \tilde{u}_\psi}{\partial Z}\bigg|_{Z=H}\right){\rm d}\rho\,, \label{force integral}
\end{align}
and evaluating this integral gives
\begin{align}
	F_x\,=\,-\frac{8\pi\eta Ru_c}{5}\log(\rho_0)+\mathcal{O}(1)\,. \label{integrated}
\end{align}
We wish to express $F_x$ in terms of $\epsilon$. As discussed in Ref.~\cite{ishikawa06, kim13}, we do not need to determine the unknown constant $D$ in order to do this, because ${\log{\rho_0}=-(1/2)\log\epsilon + \log{D}}$, so the value of $D$ can be absorbed into the $\mathcal{O}(1)$ term. Hence we obtain the expression in Eq.~\eqref{rotating squirmer force}. 

The torque $T_y$ can be calculated in the same way by integrating the differential elements of torque~\cite{ishikawa06}
\begin{align}
{\rm d}T_y=R\left[(\hat{\mathbf{n}}\cdot\hat{\mathbf{z}})\hat{\mathbf{x}}-(\hat{\mathbf{n}}\cdot\hat{\mathbf{x}})\hat{\mathbf{z}}\right]\cdot\mathbf{\upsigma}\cdot\hat{\mathbf{n}}{\rm d}S\,,
\end{align}
giving, after the same steps as above, the integral
\begin{align}
T_y\,=\,R^3\pi\int_0^{\rho_0}\frac{\eta\rho}{R}\left(\frac{\partial \tilde{u}_\rho}{\partial Z}\bigg|_{Z=H}-\frac{\partial \tilde{u}_\psi}{\partial Z}\bigg|_{Z=H}\right){\rm d}\rho\,, \label{torque}
\end{align}
which yields the expression in Eq.~\eqref{rotating squirmer torque}.

\section{Matching Lubrication and Far-field Results}

In order to obtain a result which can be compared with the simulation results everywhere, we perform a matched asymptotic expansion of the near-field and far-field results. We define $q=R/h'=1/(1+\epsilon)$. Then, the far-field corresponds to $q\rightarrow 0$, while the near-field corresponds to $\epsilon\rightarrow0$. In the near-field, the next-to-leading-order term is $\mathcal{O}(1/\log\epsilon)$, so, in order to match this term to the far-field we define the function
\begin{align}
f(q)\,=\,\frac{2q}{\log\left(\frac{1+q}{1-q}\right)}\,,
\end{align}
which has the near-field limit $f\rightarrow-2/\log(\epsilon)$, and the far-field limit $f\rightarrow1$. For intermediate values, $f(q)$ is smooth and monotonic. We then use the following matched expansion
%\begin{widetext}
\begin{align}
\frac{1}{Ru_0}\frac{{\rm d}\phi}{{\rm d}t}\,=\,\cos{\phi}q^4\left[-\frac{3}{2}+\left(\frac{21}{16}+c_1q^2\right)f(q)\right]+\beta\sin{(2\phi)}q^3\left[-\frac{3}{4}+\left(\frac{57}{64}+c_2q^2\right)f(q)\right]\,, \label{speed}
\end{align}
%\end{widetext}
where $c_1$ and $c_2$ are constants to be determined by matching to the simulations. This expansion matches both the lubrication results and the far-field results in their respective domains of applicability, with corrections of $\mathcal{O}(1/\log\epsilon)$ in the near field, and $\mathcal{O}(\beta q^5)$ and $\mathcal{O}(q^6)$ in the far-field, which is the next order of approximation there~\cite{spagnolie12}. There are two next-to-leading-order terms in the far-field because we have linearly independent contributions from the $n=1$ and $n=2$ Legendre components of the slip velocity.

With the fitting parameters, $c_1=1$, $c_2=0.29$, we obtain semi-quantitative agreement with the simulation results. Because of the very slow decay of the next-to-leading-order terms in the lubrication theory, we would not expect an exact match. Thorough testing of the lubrication theory would require simulations where the swimmer approaches much closer to the plane surface.

%\bibliography{lubricationBip}

\begin{thebibliography}{34}%
\makeatletter
\providecommand \@ifxundefined [1]{%
 \@ifx{#1\undefined}
}%
\providecommand \@ifnum [1]{%
 \ifnum #1\expandafter \@firstoftwo
 \else \expandafter \@secondoftwo
 \fi
}%
\providecommand \@ifx [1]{%
 \ifx #1\expandafter \@firstoftwo
 \else \expandafter \@secondoftwo
 \fi
}%
\providecommand \natexlab [1]{#1}%
\providecommand \enquote  [1]{``#1''}%
\providecommand \bibnamefont  [1]{#1}%
\providecommand \bibfnamefont [1]{#1}%
\providecommand \citenamefont [1]{#1}%
\providecommand \href@noop [0]{\@secondoftwo}%
\providecommand \href [0]{\begingroup \@sanitize@url \@href}%
\providecommand \@href[1]{\@@startlink{#1}\@@href}%
\providecommand \@@href[1]{\endgroup#1\@@endlink}%
\providecommand \@sanitize@url [0]{\catcode `\\12\catcode `\$12\catcode
  `\&12\catcode `\#12\catcode `\^12\catcode `\_12\catcode `\%12\relax}%
\providecommand \@@startlink[1]{}%
\providecommand \@@endlink[0]{}%
\providecommand \url  [0]{\begingroup\@sanitize@url \@url }%
\providecommand \@url [1]{\endgroup\@href {#1}{\urlprefix }}%
\providecommand \urlprefix  [0]{URL }%
\providecommand \Eprint [0]{\href }%
\providecommand \doibase [0]{http://dx.doi.org/}%
\providecommand \selectlanguage [0]{\@gobble}%
\providecommand \bibinfo  [0]{\@secondoftwo}%
\providecommand \bibfield  [0]{\@secondoftwo}%
\providecommand \translation [1]{[#1]}%
\providecommand \BibitemOpen [0]{}%
\providecommand \bibitemStop [0]{}%
\providecommand \bibitemNoStop [0]{.\EOS\space}%
\providecommand \EOS [0]{\spacefactor3000\relax}%
\providecommand \BibitemShut  [1]{\csname bibitem#1\endcsname}%
\let\auto@bib@innerbib\@empty
%</preamble>
\bibitem [{\citenamefont {{Lord Rothschild}}(1963)}]{lord63}%
  \BibitemOpen
  \bibfield  {author} {\bibinfo {author} {\bibnamefont {{Lord Rothschild}}},\
  }\href@noop {} {\bibfield  {journal} {\bibinfo  {journal} {Nature}\ }\textbf
  {\bibinfo {volume} {198}},\ \bibinfo {pages} {1221} (\bibinfo {year}
  {1963})}\BibitemShut {NoStop}%
\bibitem [{\citenamefont {Brown}\ and\ \citenamefont {Poon}(2014)}]{brown14}%
  \BibitemOpen
  \bibfield  {author} {\bibinfo {author} {\bibfnamefont {A.~T.}\ \bibnamefont
  {Brown}}\ and\ \bibinfo {author} {\bibfnamefont {W.~C.~K.}\ \bibnamefont
  {Poon}},\ }\href@noop {} {\bibfield  {journal} {\bibinfo  {journal} {Soft
  Matter}\ }\textbf {\bibinfo {volume} {10}},\ \bibinfo {pages} {4016}
  (\bibinfo {year} {2014})}\BibitemShut {NoStop}%
\bibitem [{\citenamefont {Takagi}\ \emph {et~al.}(2014)\citenamefont {Takagi},
  \citenamefont {Palacci}, \citenamefont {Braunschweig}, \citenamefont
  {Shelley},\ and\ \citenamefont {Zhang}}]{takagi14}%
  \BibitemOpen
  \bibfield  {author} {\bibinfo {author} {\bibfnamefont {D.}~\bibnamefont
  {Takagi}}, \bibinfo {author} {\bibfnamefont {J.}~\bibnamefont {Palacci}},
  \bibinfo {author} {\bibfnamefont {A.~B.}\ \bibnamefont {Braunschweig}},
  \bibinfo {author} {\bibfnamefont {M.~J.}\ \bibnamefont {Shelley}}, \ and\
  \bibinfo {author} {\bibfnamefont {J.}~\bibnamefont {Zhang}},\ }\href@noop {}
  {\bibfield  {journal} {\bibinfo  {journal} {Soft Matter}\ }\textbf {\bibinfo
  {volume} {10}},\ \bibinfo {pages} {1784} (\bibinfo {year}
  {2014})}\BibitemShut {NoStop}%
\bibitem [{\citenamefont {Brown}\ \emph {et~al.}(2016)\citenamefont {Brown},
  \citenamefont {Vladesdu}, \citenamefont {Dawson}, \citenamefont {Visser},
  \citenamefont {Schwarz-Linek}, \citenamefont {Lintuvuori},\ and\
  \citenamefont {Poon}}]{brown15}%
  \BibitemOpen
  \bibfield  {author} {\bibinfo {author} {\bibfnamefont {A.~T.}\ \bibnamefont
  {Brown}}, \bibinfo {author} {\bibfnamefont {I.~D.}\ \bibnamefont {Vladesdu}},
  \bibinfo {author} {\bibfnamefont {A.}~\bibnamefont {Dawson}}, \bibinfo
  {author} {\bibfnamefont {T.}~\bibnamefont {Visser}}, \bibinfo {author}
  {\bibfnamefont {J.}~\bibnamefont {Schwarz-Linek}}, \bibinfo {author}
  {\bibfnamefont {J.~S.}\ \bibnamefont {Lintuvuori}}, \ and\ \bibinfo {author}
  {\bibfnamefont {W.~C.~K.}\ \bibnamefont {Poon}},\ }\href@noop {} {\bibfield
  {journal} {\bibinfo  {journal} {Soft Matter}\ }\textbf {\bibinfo {volume}
  {12}},\ \bibinfo {pages} {131} (\bibinfo {year} {2016})}\BibitemShut
  {NoStop}%
\bibitem [{\citenamefont {Galadja}\ \emph {et~al.}(2007)\citenamefont
  {Galadja}, \citenamefont {Keymer}, \citenamefont {Chaikin},\ and\
  \citenamefont {Austin}}]{galadja07}%
  \BibitemOpen
  \bibfield  {author} {\bibinfo {author} {\bibfnamefont {P.}~\bibnamefont
  {Galadja}}, \bibinfo {author} {\bibfnamefont {J.}~\bibnamefont {Keymer}},
  \bibinfo {author} {\bibfnamefont {P.}~\bibnamefont {Chaikin}}, \ and\
  \bibinfo {author} {\bibfnamefont {R.}~\bibnamefont {Austin}},\ }\href@noop {}
  {\bibfield  {journal} {\bibinfo  {journal} {J. Bacteriol.}\ }\textbf
  {\bibinfo {volume} {189}},\ \bibinfo {pages} {8704} (\bibinfo {year}
  {2007})}\BibitemShut {NoStop}%
\bibitem [{\citenamefont {Berke}\ \emph {et~al.}(2008)\citenamefont {Berke},
  \citenamefont {Turner}, \citenamefont {Berg},\ and\ \citenamefont
  {Lauga}}]{berke08}%
  \BibitemOpen
  \bibfield  {author} {\bibinfo {author} {\bibfnamefont {A.~P.}\ \bibnamefont
  {Berke}}, \bibinfo {author} {\bibfnamefont {L.}~\bibnamefont {Turner}},
  \bibinfo {author} {\bibfnamefont {H.~C.}\ \bibnamefont {Berg}}, \ and\
  \bibinfo {author} {\bibfnamefont {E.}~\bibnamefont {Lauga}},\ }\href@noop {}
  {\bibfield  {journal} {\bibinfo  {journal} {Phys Rev Lett}\ }\textbf
  {\bibinfo {volume} {101}},\ \bibinfo {pages} {038102} (\bibinfo {year}
  {2008})}\BibitemShut {NoStop}%
\bibitem [{\citenamefont {Li}\ and\ \citenamefont {Tang}(2009)}]{li09}%
  \BibitemOpen
  \bibfield  {author} {\bibinfo {author} {\bibfnamefont {G.}~\bibnamefont
  {Li}}\ and\ \bibinfo {author} {\bibfnamefont {J.~X.}\ \bibnamefont {Tang}},\
  }\href@noop {} {\bibfield  {journal} {\bibinfo  {journal} {Phys Rev Lett}\
  }\textbf {\bibinfo {volume} {103}},\ \bibinfo {pages} {78101} (\bibinfo
  {year} {2009})}\BibitemShut {NoStop}%
\bibitem [{\citenamefont {Elgeti}\ and\ \citenamefont
  {Gompper}(2013)}]{elgeti13}%
  \BibitemOpen
  \bibfield  {author} {\bibinfo {author} {\bibfnamefont {J.}~\bibnamefont
  {Elgeti}}\ and\ \bibinfo {author} {\bibfnamefont {G.}~\bibnamefont
  {Gompper}},\ }\href@noop {} {\bibfield  {journal} {\bibinfo  {journal} {EPL}\
  }\textbf {\bibinfo {volume} {101}},\ \bibinfo {pages} {48003} (\bibinfo
  {year} {2013})}\BibitemShut {NoStop}%
\bibitem [{\citenamefont {Uspal}\ \emph {et~al.}(2015)\citenamefont {Uspal},
  \citenamefont {Popescu}, \citenamefont {Dietrich},\ and\ \citenamefont
  {Tasinkevych}}]{uspal15}%
  \BibitemOpen
  \bibfield  {author} {\bibinfo {author} {\bibfnamefont {W.~E.}\ \bibnamefont
  {Uspal}}, \bibinfo {author} {\bibfnamefont {M.~N.}\ \bibnamefont {Popescu}},
  \bibinfo {author} {\bibfnamefont {S.}~\bibnamefont {Dietrich}}, \ and\
  \bibinfo {author} {\bibfnamefont {M.}~\bibnamefont {Tasinkevych}},\
  }\href@noop {} {\bibfield  {journal} {\bibinfo  {journal} {Soft Matter}\
  }\textbf {\bibinfo {volume} {11}},\ \bibinfo {pages} {434} (\bibinfo {year}
  {2015})}\BibitemShut {NoStop}%
\bibitem [{\citenamefont {Schaar}\ \emph {et~al.}(2015)\citenamefont {Schaar},
  \citenamefont {Z{\"o}ttl},\ and\ \citenamefont {Stark}}]{schaar15}%
  \BibitemOpen
  \bibfield  {author} {\bibinfo {author} {\bibfnamefont {K.}~\bibnamefont
  {Schaar}}, \bibinfo {author} {\bibfnamefont {A.}~\bibnamefont {Z{\"o}ttl}}, \
  and\ \bibinfo {author} {\bibfnamefont {H.}~\bibnamefont {Stark}},\
  }\href@noop {} {\bibfield  {journal} {\bibinfo  {journal} {Phys. Rev. Lett.}\
  }\textbf {\bibinfo {volume} {115}},\ \bibinfo {pages} {038101} (\bibinfo
  {year} {2015})}\BibitemShut {NoStop}%
\bibitem [{\citenamefont {Ishimoto}\ and\ \citenamefont
  {Gaffney}(2013)}]{ishimoto13}%
  \BibitemOpen
  \bibfield  {author} {\bibinfo {author} {\bibfnamefont {K.}~\bibnamefont
  {Ishimoto}}\ and\ \bibinfo {author} {\bibfnamefont {E.~A.}\ \bibnamefont
  {Gaffney}},\ }\href@noop {} {\bibfield  {journal} {\bibinfo  {journal}
  {Physical Review E}\ }\textbf {\bibinfo {volume} {88}},\ \bibinfo {pages}
  {062702} (\bibinfo {year} {2013})}\BibitemShut {NoStop}%
\bibitem [{\citenamefont {Li}\ and\ \citenamefont {Ardekani}(2014)}]{li14}%
  \BibitemOpen
  \bibfield  {author} {\bibinfo {author} {\bibfnamefont {G.-J.}\ \bibnamefont
  {Li}}\ and\ \bibinfo {author} {\bibfnamefont {A.~M.}\ \bibnamefont
  {Ardekani}},\ }\href@noop {} {\bibfield  {journal} {\bibinfo  {journal}
  {Phys. Rev. E}\ }\textbf {\bibinfo {volume} {90}},\ \bibinfo {pages} {013010}
  (\bibinfo {year} {2014})}\BibitemShut {NoStop}%
\bibitem [{\citenamefont {Lighthill}(1952)}]{lighthill52}%
  \BibitemOpen
  \bibfield  {author} {\bibinfo {author} {\bibfnamefont {M.~J.}\ \bibnamefont
  {Lighthill}},\ }\href@noop {} {\bibfield  {journal} {\bibinfo  {journal}
  {Communications on Pure and Applied Mathematics}\ }\textbf {\bibinfo {volume}
  {5}},\ \bibinfo {pages} {109} (\bibinfo {year} {1952})}\BibitemShut {NoStop}%
\bibitem [{\citenamefont {Z\"ottl}\ and\ \citenamefont
  {Stark}(2014)}]{zottl14}%
  \BibitemOpen
  \bibfield  {author} {\bibinfo {author} {\bibfnamefont {A.}~\bibnamefont
  {Z\"ottl}}\ and\ \bibinfo {author} {\bibfnamefont {H.}~\bibnamefont
  {Stark}},\ }\href@noop {} {\bibfield  {journal} {\bibinfo  {journal} {Phys.
  Rev. Lett.}\ }\textbf {\bibinfo {volume} {112}},\ \bibinfo {pages} {118101}
  (\bibinfo {year} {2014})}\BibitemShut {NoStop}%
\bibitem [{\citenamefont {Lambert}\ \emph {et~al.}(2013)\citenamefont
  {Lambert}, \citenamefont {Picano}, \citenamefont {Brandt},\ and\
  \citenamefont {Breugem}}]{Lambert13}%
  \BibitemOpen
  \bibfield  {author} {\bibinfo {author} {\bibfnamefont {R.~A.}\ \bibnamefont
  {Lambert}}, \bibinfo {author} {\bibfnamefont {F.}~\bibnamefont {Picano}},
  \bibinfo {author} {\bibfnamefont {L.}~\bibnamefont {Brandt}}, \ and\ \bibinfo
  {author} {\bibfnamefont {W.~P.}\ \bibnamefont {Breugem}},\ }\href@noop {}
  {\bibfield  {journal} {\bibinfo  {journal} {J. FLuid. Mech.}\ }\textbf
  {\bibinfo {volume} {733}},\ \bibinfo {pages} {528} (\bibinfo {year}
  {2013})}\BibitemShut {NoStop}%
\bibitem [{\citenamefont {Cates}\ \emph {et~al.}(2004)\citenamefont {Cates},
  \citenamefont {Stratford}, \citenamefont {Adhikari}, \citenamefont
  {Stansell}, \citenamefont {Desplat}, \citenamefont {Pagonabarraga},\ and\
  \citenamefont {Wagner}}]{mikeLB}%
  \BibitemOpen
  \bibfield  {author} {\bibinfo {author} {\bibfnamefont {M.~E.}\ \bibnamefont
  {Cates}}, \bibinfo {author} {\bibfnamefont {K.}~\bibnamefont {Stratford}},
  \bibinfo {author} {\bibfnamefont {R.}~\bibnamefont {Adhikari}}, \bibinfo
  {author} {\bibfnamefont {P.}~\bibnamefont {Stansell}}, \bibinfo {author}
  {\bibfnamefont {J.-C.}\ \bibnamefont {Desplat}}, \bibinfo {author}
  {\bibfnamefont {I.}~\bibnamefont {Pagonabarraga}}, \ and\ \bibinfo {author}
  {\bibfnamefont {A.~J.}\ \bibnamefont {Wagner}},\ }\href@noop {} {\bibfield
  {journal} {\bibinfo  {journal} {J. Phys. Condens. Mater.}\ }\textbf {\bibinfo
  {volume} {16}},\ \bibinfo {pages} {S3903} (\bibinfo {year}
  {2004})}\BibitemShut {NoStop}%
\bibitem [{\citenamefont {Magar}\ \emph {et~al.}(2003)\citenamefont {Magar},
  \citenamefont {Goto},\ and\ \citenamefont {Pedley}}]{Magar03}%
  \BibitemOpen
  \bibfield  {author} {\bibinfo {author} {\bibfnamefont {V.}~\bibnamefont
  {Magar}}, \bibinfo {author} {\bibfnamefont {T.}~\bibnamefont {Goto}}, \ and\
  \bibinfo {author} {\bibfnamefont {T.~J.}\ \bibnamefont {Pedley}},\
  }\href@noop {} {\bibfield  {journal} {\bibinfo  {journal} {Quart. J. Mech.
  Appl. Math.}\ }\textbf {\bibinfo {volume} {56}},\ \bibinfo {pages} {65}
  (\bibinfo {year} {2003})}\BibitemShut {NoStop}%
\bibitem [{\citenamefont {Ladd}(1994{\natexlab{a}})}]{ladd1}%
  \BibitemOpen
  \bibfield  {author} {\bibinfo {author} {\bibfnamefont {A.~J.~C.}\
  \bibnamefont {Ladd}},\ }\href@noop {} {\bibfield  {journal} {\bibinfo
  {journal} {J. Fluid Mech.}\ }\textbf {\bibinfo {volume} {271}},\ \bibinfo
  {pages} {285} (\bibinfo {year} {1994}{\natexlab{a}})}\BibitemShut {NoStop}%
\bibitem [{\citenamefont {Ladd}(1994{\natexlab{b}})}]{ladd2}%
  \BibitemOpen
  \bibfield  {author} {\bibinfo {author} {\bibfnamefont {A.~J.~C.}\
  \bibnamefont {Ladd}},\ }\href@noop {} {\bibfield  {journal} {\bibinfo
  {journal} {J. Fluid Mech.}\ }\textbf {\bibinfo {volume} {271}},\ \bibinfo
  {pages} {311} (\bibinfo {year} {1994}{\natexlab{b}})}\BibitemShut {NoStop}%
\bibitem [{\citenamefont {Nguyen}\ and\ \citenamefont {Ladd}(2002)}]{ladd3}%
  \BibitemOpen
  \bibfield  {author} {\bibinfo {author} {\bibfnamefont {N.-Q.}\ \bibnamefont
  {Nguyen}}\ and\ \bibinfo {author} {\bibfnamefont {A.~J.~C.}\ \bibnamefont
  {Ladd}},\ }\href@noop {} {\bibfield  {journal} {\bibinfo  {journal} {Phys.
  Rev. E}\ }\textbf {\bibinfo {volume} {66}},\ \bibinfo {pages} {046708}
  (\bibinfo {year} {2002})}\BibitemShut {NoStop}%
\bibitem [{\citenamefont {Llopis}\ and\ \citenamefont
  {Pagonabarraga}(2010)}]{ignacio1}%
  \BibitemOpen
  \bibfield  {author} {\bibinfo {author} {\bibfnamefont {I.}~\bibnamefont
  {Llopis}}\ and\ \bibinfo {author} {\bibfnamefont {I.}~\bibnamefont
  {Pagonabarraga}},\ }\href@noop {} {\bibfield  {journal} {\bibinfo  {journal}
  {J. Non-Newtonian Fluid Mech.}\ }\textbf {\bibinfo {volume} {165}},\ \bibinfo
  {pages} {946} (\bibinfo {year} {2010})}\BibitemShut {NoStop}%
\bibitem [{\citenamefont {Pagonabarraga}\ and\ \citenamefont
  {Llopis}(2013)}]{ignacio2}%
  \BibitemOpen
  \bibfield  {author} {\bibinfo {author} {\bibfnamefont {I.}~\bibnamefont
  {Pagonabarraga}}\ and\ \bibinfo {author} {\bibfnamefont {I.}~\bibnamefont
  {Llopis}},\ }\href@noop {} {\bibfield  {journal} {\bibinfo  {journal} {Soft
  Matter}\ }\textbf {\bibinfo {volume} {9}},\ \bibinfo {pages} {7174} (\bibinfo
  {year} {2013})}\BibitemShut {NoStop}%
\bibitem [{\citenamefont {Adhikari}\ \emph {et~al.}(2005)\citenamefont
  {Adhikari}, \citenamefont {Stratford}, \citenamefont {Cates},\ and\
  \citenamefont {Wagner}}]{adhikari}%
  \BibitemOpen
  \bibfield  {author} {\bibinfo {author} {\bibfnamefont {R.}~\bibnamefont
  {Adhikari}}, \bibinfo {author} {\bibfnamefont {K.}~\bibnamefont {Stratford}},
  \bibinfo {author} {\bibfnamefont {M.~E.}\ \bibnamefont {Cates}}, \ and\
  \bibinfo {author} {\bibfnamefont {A.~J.}\ \bibnamefont {Wagner}},\
  }\href@noop {} {\bibfield  {journal} {\bibinfo  {journal} {Europhys. Lett.}\
  }\textbf {\bibinfo {volume} {71}},\ \bibinfo {pages} {473} (\bibinfo {year}
  {2005})}\BibitemShut {NoStop}%
\bibitem [{Sup()}]{Suppl}%
  \BibitemOpen
  \href@noop {} {\ }\bibinfo {note} {See Supplementary Information online XXX,
  for additional figures S1-S3, additional information on the model, far-field
  equations $\tfrac{\mathrm{d}x}{\mathrm{d}t}$ and
  $\tfrac{\mathrm{d}z}{\mathrm{d}t}$, and for details of the derivation the
  near-field equations and of how the far and near-field contributions can be
  combined.}\BibitemShut {Stop}%
\bibitem [{Note1()}]{Note1}%
  \BibitemOpen
  \bibinfo {note} {These would correspond to $R\sim $ 10 $\mu $m, $u_0\sim
  10\protect \genfrac {}{}{}1{\protect \mathrm {mm}}{\protect \mathrm {s}}$,
  and a distance between walls of $\sim 100\mu $m.}\BibitemShut {Stop}%
\bibitem [{\citenamefont {Das}\ \emph {et~al.}(2015)\citenamefont {Das},
  \citenamefont {Garg}, \citenamefont {Campbell}, \citenamefont {Howse},
  \citenamefont {Sen}, \citenamefont {Velegol}, \citenamefont {Golestanian},\
  and\ \citenamefont {Ebbens}}]{das15}%
  \BibitemOpen
  \bibfield  {author} {\bibinfo {author} {\bibfnamefont {S.}~\bibnamefont
  {Das}}, \bibinfo {author} {\bibfnamefont {A.}~\bibnamefont {Garg}}, \bibinfo
  {author} {\bibfnamefont {A.~I.}\ \bibnamefont {Campbell}}, \bibinfo {author}
  {\bibfnamefont {J.}~\bibnamefont {Howse}}, \bibinfo {author} {\bibfnamefont
  {A.}~\bibnamefont {Sen}}, \bibinfo {author} {\bibfnamefont {D.}~\bibnamefont
  {Velegol}}, \bibinfo {author} {\bibfnamefont {R.}~\bibnamefont
  {Golestanian}}, \ and\ \bibinfo {author} {\bibfnamefont {S.~J.}\ \bibnamefont
  {Ebbens}},\ }\href@noop {} {\bibfield  {journal} {\bibinfo  {journal} {Nat.
  Commun.}\ }\textbf {\bibinfo {volume} {6}} (\bibinfo {year}
  {2015})}\BibitemShut {NoStop}%
\bibitem [{\citenamefont {Spagnolie}\ and\ \citenamefont
  {Lauga}(2012)}]{spagnolie12}%
  \BibitemOpen
  \bibfield  {author} {\bibinfo {author} {\bibfnamefont {S.~E.}\ \bibnamefont
  {Spagnolie}}\ and\ \bibinfo {author} {\bibfnamefont {E.}~\bibnamefont
  {Lauga}},\ }\href@noop {} {\bibfield  {journal} {\bibinfo  {journal} {J.Fluid
  Mech.}\ }\textbf {\bibinfo {volume} {700}},\ \bibinfo {pages} {105} (\bibinfo
  {year} {2012})}\BibitemShut {NoStop}%
\bibitem [{\citenamefont {Z{\"o}ttl}(2014)}]{ZottlThesis}%
  \BibitemOpen
  \bibfield  {author} {\bibinfo {author} {\bibfnamefont {A.}~\bibnamefont
  {Z{\"o}ttl}},\ }\emph {\bibinfo {title} {Hydrodynamics of Microswimmers in
  Confinement and in Poiseuille Flow}},\ \href@noop {} {Ph.D. thesis},\
  \bibinfo  {school} {Technischen Universit{\"a}t Berlin} (\bibinfo {year}
  {2014})\BibitemShut {NoStop}%
\bibitem [{\citenamefont {Cichocki}\ and\ \citenamefont
  {Jones}(1998)}]{cichocki98}%
  \BibitemOpen
  \bibfield  {author} {\bibinfo {author} {\bibfnamefont {B.}~\bibnamefont
  {Cichocki}}\ and\ \bibinfo {author} {\bibfnamefont {R.}~\bibnamefont
  {Jones}},\ }\href@noop {} {\bibfield  {journal} {\bibinfo  {journal} {Phys.
  Rev. Lett.}\ }\textbf {\bibinfo {volume} {258}},\ \bibinfo {pages} {273}
  (\bibinfo {year} {1998})}\BibitemShut {NoStop}%
\bibitem [{\citenamefont {Ledesma-Aguilar}\ and\ \citenamefont
  {Yeomans}(2013)}]{ledesma13}%
  \BibitemOpen
  \bibfield  {author} {\bibinfo {author} {\bibfnamefont {R.}~\bibnamefont
  {Ledesma-Aguilar}}\ and\ \bibinfo {author} {\bibfnamefont {J.~M.}\
  \bibnamefont {Yeomans}},\ }\href@noop {} {\bibfield  {journal} {\bibinfo
  {journal} {Phys. Rev. Lett.}\ }\textbf {\bibinfo {volume} {111}},\ \bibinfo
  {pages} {138101} (\bibinfo {year} {2013})}\BibitemShut {NoStop}%
\bibitem [{\citenamefont {Wang}\ \emph {et~al.}(2015)\citenamefont {Wang},
  \citenamefont {In}, \citenamefont {Blanc}, \citenamefont {Nobili},\ and\
  \citenamefont {Stocco}}]{wang15}%
  \BibitemOpen
  \bibfield  {author} {\bibinfo {author} {\bibfnamefont {X.}~\bibnamefont
  {Wang}}, \bibinfo {author} {\bibfnamefont {M.}~\bibnamefont {In}}, \bibinfo
  {author} {\bibfnamefont {C.}~\bibnamefont {Blanc}}, \bibinfo {author}
  {\bibfnamefont {M.}~\bibnamefont {Nobili}}, \ and\ \bibinfo {author}
  {\bibfnamefont {A.}~\bibnamefont {Stocco}},\ }\href {\doibase DOI:
  10.1039/C5SM01111F} {\bibfield  {journal} {\bibinfo  {journal} {Soft Matter}\
  }\textbf {\bibinfo {volume} {12}},\ \bibinfo {pages} {7376} (\bibinfo {year}
  {2015})}\BibitemShut {NoStop}%
\bibitem [{\citenamefont {Theurkauff}\ \emph {et~al.}(2012)\citenamefont
  {Theurkauff}, \citenamefont {Cottin-Bizonne}, \citenamefont {Palacci},
  \citenamefont {Ybert},\ and\ \citenamefont {Bocquet}}]{theurkauff12}%
  \BibitemOpen
  \bibfield  {author} {\bibinfo {author} {\bibfnamefont {I.}~\bibnamefont
  {Theurkauff}}, \bibinfo {author} {\bibfnamefont {C.}~\bibnamefont
  {Cottin-Bizonne}}, \bibinfo {author} {\bibfnamefont {J.}~\bibnamefont
  {Palacci}}, \bibinfo {author} {\bibfnamefont {C.}~\bibnamefont {Ybert}}, \
  and\ \bibinfo {author} {\bibfnamefont {L.}~\bibnamefont {Bocquet}},\
  }\href@noop {} {\bibfield  {journal} {\bibinfo  {journal} {Phys. Rev. Lett.}\
  }\textbf {\bibinfo {volume} {108}},\ \bibinfo {pages} {268303} (\bibinfo
  {year} {2012})}\BibitemShut {NoStop}%
\bibitem [{\citenamefont {Bickel}\ \emph {et~al.}(2013)\citenamefont {Bickel},
  \citenamefont {Majee},\ and\ \citenamefont {W{\"u}rger}}]{bickel13}%
  \BibitemOpen
  \bibfield  {author} {\bibinfo {author} {\bibfnamefont {T.}~\bibnamefont
  {Bickel}}, \bibinfo {author} {\bibfnamefont {A.}~\bibnamefont {Majee}}, \
  and\ \bibinfo {author} {\bibfnamefont {A.}~\bibnamefont {W{\"u}rger}},\
  }\href@noop {} {\bibfield  {journal} {\bibinfo  {journal} {Physical Review
  E}\ }\textbf {\bibinfo {volume} {88}},\ \bibinfo {pages} {012301} (\bibinfo
  {year} {2013})}\BibitemShut {NoStop}%
\bibitem [{\citenamefont {Ginot}\ \emph {et~al.}(2015)\citenamefont {Ginot},
  \citenamefont {Theurkauff}, \citenamefont {Levis}, \citenamefont {Ybert},
  \citenamefont {Bocquet}, \citenamefont {Berthier},\ and\ \citenamefont
  {Cottin-Bizonne}}]{ginot15}%
  \BibitemOpen
  \bibfield  {author} {\bibinfo {author} {\bibfnamefont {F.}~\bibnamefont
  {Ginot}}, \bibinfo {author} {\bibfnamefont {I.}~\bibnamefont {Theurkauff}},
  \bibinfo {author} {\bibfnamefont {D.}~\bibnamefont {Levis}}, \bibinfo
  {author} {\bibfnamefont {C.}~\bibnamefont {Ybert}}, \bibinfo {author}
  {\bibfnamefont {L.}~\bibnamefont {Bocquet}}, \bibinfo {author} {\bibfnamefont
  {L.}~\bibnamefont {Berthier}}, \ and\ \bibinfo {author} {\bibfnamefont
  {C.}~\bibnamefont {Cottin-Bizonne}},\ }\href@noop {} {\bibfield  {journal}
  {\bibinfo  {journal} {Phys. Rev. X}\ }\textbf {\bibinfo {volume} {5}},\
  \bibinfo {pages} {011004} (\bibinfo {year} {2015})}\BibitemShut {NoStop}%
\end{thebibliography}

\begin{thebibliography}{3}
\expandafter\ifx\csname natexlab\endcsname\relax\def\natexlab#1{#1}\fi
\expandafter\ifx\csname bibnamefont\endcsname\relax
  \def\bibnamefont#1{#1}\fi
\expandafter\ifx\csname bibfnamefont\endcsname\relax
  \def\bibfnamefont#1{#1}\fi
\expandafter\ifx\csname citenamefont\endcsname\relax
  \def\citenamefont#1{#1}\fi
\expandafter\ifx\csname url\endcsname\relax
  \def\url#1{\texttt{#1}}\fi
\expandafter\ifx\csname urlprefix\endcsname\relax\def\urlprefix{URL }\fi
\providecommand{\bibinfo}[2]{#2}
\providecommand{\eprint}[2][]{\url{#2}}

\bibitem[{\citenamefont{Spagnolie and Lauga}(2012)}]{spagnolie12}
\bibinfo{author}{\bibfnamefont{S.~E.} \bibnamefont{Spagnolie}}
  \bibnamefont{and} \bibinfo{author}{\bibfnamefont{E.}~\bibnamefont{Lauga}},
  \bibinfo{journal}{J. Fluid Mech.} \textbf{\bibinfo{volume}{700}},
  \bibinfo{pages}{105} (\bibinfo{year}{2012}).

\bibitem[{\citenamefont{Ishikawa et~al.}(2006)\citenamefont{Ishikawa, Simmonds,
  and Pedley}}]{ishikawa06}
\bibinfo{author}{\bibfnamefont{T.}~\bibnamefont{Ishikawa}},
  \bibinfo{author}{\bibfnamefont{M.~P.} \bibnamefont{Simmonds}},
  \bibnamefont{and} \bibinfo{author}{\bibfnamefont{T.~J.}
  \bibnamefont{Pedley}}, \bibinfo{journal}{J. Fluid Mech.}
  \textbf{\bibinfo{volume}{568}}, \bibinfo{pages}{119} (\bibinfo{year}{2006}).

\bibitem[{\citenamefont{Kim and Karrila}(2013)}]{kim13}
\bibinfo{author}{\bibfnamefont{S.}~\bibnamefont{Kim}} \bibnamefont{and}
  \bibinfo{author}{\bibfnamefont{S.~J.} \bibnamefont{Karrila}},
  \emph{\bibinfo{title}{Microhydrodynamics: principles and selected
  applications}} (\bibinfo{publisher}{Courier Dover Publications},
  \bibinfo{year}{2013}).

\end{thebibliography}

\end{document}